\documentclass[12pt]{article}
 \usepackage[dvips]{graphicx}
 \usepackage[dvips]{graphics}
 \usepackage{amsmath}
 \usepackage{enumerate}
 \usepackage{psfrag}
 \newcommand{\newc}{\newcommand}
 \newc{\ra}{\rightarrow}
 \newc{\lra}{\leftrightarrow}
 \newc{\beq}{\begin{equation}}
 \newc{\eeq}{\end{equation}}
 \newc{\barr}{\begin{eqnarray}}
 \newc{\barra}{\begin{eqnarray*}}
 \newc{\earr}{\end{eqnarray}}
 \newc{\earra}{\end{eqnarray*}}
 \newc{\texa}{\textstyle}
 \newc{\paral}{\parallel}
 \newc{\und}{\underline}
 \newc{\pars}{\partial}
 \newc{\nonu}{\nonumber \\}
 \newc{\nln}{\\ \vspace{2mm}}
 \newc{\oms}{\omega_{{\bf E}\times {\bf B}}}

\begin{document}

 \begin{center}
 {\LARGE \bf  Two-fluid tokamak equilibria  with \\
 reversed magnetic shear and sheared flow
 \footnote{A preliminary version of this study was presented in the 10th
 European Fusion Theory Conference (Helsinki, Finland, 8-10 September
 2003).}}

 \vspace{2mm}

 {\large G. Poulipoulis$^\dag$\footnote{me00584@cc.uoi.gr},
 G. N. Throumoulopoulos$^\dag$\footnote{gthroum@cc.uoi.gr}, H.
 Tasso$^\star$\footnote{het@ipp.mpg.de}}

 \vspace*{1mm}

 $^\dag${\it University of Ioannina, Association Euratom - Hellenic
 Republic,\\ \vspace{-1mm}
 Section of Theoretical Physics, GR 451 10 Ioannina, Greece}

 \vspace{1mm} \noindent $^\star${\it  Max-Planck-Institut f\"{u}r
 Plasmaphysik, Euratom Association,\\ \vspace{-1mm}
 D-85748 Garching, Germany }
 \end{center}

 \begin{center}
 {\bf Abstract}
 \end{center}
\noindent
 The aim of the present work is to investigate tokamak
 equilibria with reversed magnetic shear and sheared flow, which may
 play a role in the formation of internal transport barriers (ITBs),
 within the framework of two-fluid model. The study is based on exact
 self-consistent solutions in cylindrical geometry by means of which
 the impact of the magnetic shear, $s$, and the ``toroidal" (axial) and
 ``poloidal" (azimuthal) ion velocity components, $v_{iz}$ and
 $v_{i\theta}$, on the radial electric field, $E_r$, its shear,
 $|dE_r/dr|$, and the shear of the   ${\bf E}\times {\bf B}$
 velocity, $\oms\equiv|d/dr({\bf E}\times {\bf B}/B^2)|$, is
 examined.  For a wide parametric regime of experimental concern it
 turns out that the contributions of the $v_{iz}$, $v_{i\theta}$ and
 pressure gradient ($\nabla P_i$) terms to $E_r$, $|E_r'|$ and $\oms$ are of the same order of
 magnitude.  The contribution of the $\nabla P_i$ term  is
 missing in the framework of magnetohydrodynamics (MHD) [G.
 Poulipoulis et al. Plasma Phys. Control. Fusion {\bf 46} (2004)
 639]. The impact of $s$ on $\oms$ through the $\nabla P_i$ term
 is stronger than that through the velocity terms; in particular for
 $B_z=$ constant, the contributions of the $\nabla P_i $ and velocity
 terms to $\oms$ at the point where $dE_r/dr=0$ are proportional to
 $(1-s)(2-s)$ and $(1-s)$, respectively.  The results indicate that,
 alike MHD,  the magnetic shear and the sheared toroidal and poloidal
 velocities  act synergetically in producing electric fields and
 therefore $\oms$ profiles compatible with ones observed in
 discharges with ITBs; owing to the $\nabla
 P_i$ term, however, the impact of $s$ on $E_r$, $|E_r'|$ and $\oms$
 is stronger than that in MHD.
 \vspace{0.4cm} \\

 \newpage
 \noindent
 {\bf \large Introduction}\\

 Tokamak discharges with improved energy and particle confinement
 properties in connection with internal transport barriers (ITBs)
 have certain attractive features, such as a large bootstrap current
 fraction, which suggest a potential route to steady-state mode of
 operation desirable for fusion power plants.  Long quasi-steady or
 steady ITB states have been obtained in different tokamaks, e.g ASDEX Upgrate \cite{Gr,Zo}, JT-60U
 \cite{Id}, Tore-Supra \cite{Li1}, and JET \cite{Li2} where ITBs were
 maintained for up to 11 s. The ITBs usually are associated with
 reversed magnetic shear profiles \cite{str},\cite{con} and their main
 characteristics are steep pressure profiles in the barrier region
 \cite{lev} and radial electric fields associated with sheared flows
 \cite{tal,CaWa}. The mechanism responsible for the formation of ITBs
 and the underlying physics is  not completely understood. Most
 theoretical models supported by experimental observations rely on
 suppression of microinstability induced transport  in connection
 with reversed magnetic shear, $s<0$, sheared  flow, the radial electric
 field, $E_r$, its shear, $|E_r^\prime|$, and most importantly the
 ${\bf E}\times {\bf B}$ velocity shear,
 \beq
 \omega_{{\bf E}\times {\bf B}}=\left
 |\frac{d}{dr} \frac{{\bf E}\times {\bf B}}{B^2}\right|.
 \label{1}
 \eeq
 In particular  the ${\bf E}\times{\bf B}$ velocity shear may lead
 to a reduction in the amplitude of turbulent fluctuations, even to
 their suppression, or to a decrease in the radial correlation lengths
 \cite{Conn}. Although there are experimental observations supporting
 this scenario, the overall experimental evidence  up to date is rather
 complicated, not universal in the various tokamak machines and has not
 made clear whether the  magnetic shear
 or the sheared flow (toroidal or poloidal) are more important for the ITB
 formation. A discussion on this issue is made in the Introduction of Ref.
 \cite{Pou}. Also, the experimental and theoretical knowledge on discharges
 with ITBs was reviewed recently in Refs. \cite{Conn} and \cite{Wo1}.

 In a previous work \cite{Pou} we studied magnetohydrodynamic (MHD)
 equilibrium states with reversed magnetic shear and sheared flow in
 cylindrical geometry. In particular, presuming that $E_r$, $E_r^\prime$ and
 $\omega_{{\bf E}\times {\bf B}}$ are of relevance to the formation of ITBs
 we examined how these quantities are affected by the  magnetic shear and
 sheared flow  and found that the latter quantities act synergetically in
 increasing $\omega_{{\bf E}\times {\bf B}}$ with the impact of the flow,
 in particular the poloidal one, being stronger than that of the magnetic
 shear, $s$. The present work aims at extending the study to the framework
 of the two fluid model. This  model is advantageous over MHD in that the
 contribution of the ion pressure gradient ($\nabla P_i$) term  to  $E_r$,
 contribution which is missing in MHD, can be obtained from the ion (or electron) momentum
 equation. In addition the current density can be expressed self-consistently
 in terms of the ion and electron fluid velocities. Also, we shall examine the
 impact of certain local characteristics of the safety factor profile, i.e.
 the minimum of $q$ and its position,  on the aforementioned quantities (not
 addressed in Ref. \cite{Pou}) and the relative sign of the ``toroidal" (axial) ion
 velocity, $v_{iz}$, ``poloidal" (azimuthal) ion velocity, $v_{i\theta}$,  and toroidal
 magnetic field, $B_z$. It turns out that, owing to the $\nabla P_i$ term,
 the impact of $s$ on  $E_r$, $E_r^\prime$ and  $\oms$  is stronger than
 that in MHD. The contribution of both flow components, however, remains
 significant. In addition in many cases $s$ enhances the velocity contribution
 to these quantities which, alike in MHD, indicates a synergism of $s$ and the flow.

 The work will be conducted through the following steps. Exact solutions
 of a slightly reduced set of two-fluid equilibrium equations for a cylindrical
 magnetically confined plasma are constructed in section 2 by prescribing the
 profiles of certain free quantities, including the safety factor and the
 toroidal and poloidal ion velocities,
 in accord with ITB experimental ones. Then in section 3 we examine the
 impact of $s$, the velocity, the velocity shear, the local characteristics
 of $q$, and the relative signs of the velocity components and $B_z$
 on $E_r$, $E_r^\prime$ and $\oms$. The characteristics of the pressure
 and toroidal current density are also briefly discussed. The conclusions
 are summarized in section 4. \vspace{4mm} \\

 {\bf 2. Two-fluid cylindrical equilibria with reversed magnetic shear}\\

 The two-fluid equilibrium states of an ideal quasineutral plasma are governed by
 the following set of equations written
 in
 Gaussian units with both $4\pi$ and the velocity of light being set
 to unity:
 \begin{eqnarray}
 {\bf \nabla}\cdot (n{\bf v}_\alpha)=0, \label{2} \\
 m_\alpha n_\alpha({\bf v}_\alpha\cdot{\bf \nabla}){\bf v}_\alpha=-{\bf
 \nabla}P_\alpha+q_\alpha n_\alpha({\bf E}+{\bf v}_\alpha\times
 {\bf B}), \label{3} \\
 {\bf v}_\alpha\cdot {\bf \nabla }T_\alpha =0, \label{3a} \\
 Z_in_i\approx n_e=n, \label{5} \\
 {\bf \nabla}\times{\bf E}=0, \label{6} \\
 {\bf \nabla}\cdot{\bf B}=0, \label{7} \\
 {\bf \nabla}\times{\bf B}=\sum_\alpha{n_\alpha q_\alpha{\bf
 v}_\alpha}={\bf J}, \label{8}
 \end{eqnarray}
 where the index $\alpha$  denotes the particle species ($\alpha=i$ for
 ions and $e$ for electrons); $n$ is the plasma density  in connection
 with the quasi-neutrality condition (\ref{5}); $q_\alpha$ is the charge
 of each particle species with $Z_i$ being the  atomic number. The
 rest of the notation is standard. The energy equation (\ref{3a}),
 associated with the fact that for fusion plasmas the heat conduction
 along ${\bf B}$ is large and therefore the temperature becomes uniform on
 magnetic surfaces on a fast time scale, is particularly appropriate
 for  electrons. For  ions one alternatively can use an adiabatic
 energy equation:
 \begin{equation}
 {\bf v}_i\cdot{\bf \nabla}P_i+\gamma P_i{\bf \nabla}
 \cdot{\bf v}_i=0.
 \label{4}
 \end{equation}
 Compared with the respective set of MHD equations (see for example
 Eqs. (2)-(6) of Ref. \cite{Pou}) Eqs. (\ref{2}-\ref{8}) are
 advantageous in two respects: (i) the momentum equation includes the
 electric field and therefore the pressure gradient contribution to
 $\bf E$ can be calculated from this equation; this contribution is missing
 in the frame of MHD because $\bf E$ is calculated by Ohm's law,
  ${\bf E} + {\bf v}\times {\bf B} =0$,
 and (ii) the current density ${\bf J}$ is related self-consistently
 to the fluid species velocities [Eq. (\ref{8})].

 The system under consideration is a cylindrical plasma of circular
 cross-section confined by a magnetic field having toroidal and poloidal
 components $B_z$ and $B_\theta$ respectively. Also the velocity has
  toroidal and  poloidal  components and the electric field is
 radial. Because of  symmetry any equilibrium quantity depends
 solely on the radial distance $r$; therefore Eqs. (\ref{2}),
 (\ref{3a})[and (\ref{4})], (\ref{6}), and (\ref{7}) are  identically satisfied.
 Also the flow for both fluid species is incompressible ($\nabla \cdot {\bf v}_\alpha
 =0$). Under these considerations 6 out of the 12 scalar quantities
 remain free and can be prescribed.

 Adding Eq. (\ref{4}) for  ions and electrons yields the MHD
 momentum equation
 \beq
 \frac{d}{dr}\left(P + \frac{B_\theta^2 + B_z^2}{2}\right)+
 \left(1-M_\theta^2\right)\frac{B_\theta^2}{r} =0,
 \label{4a}
 \eeq
 where
 $$
 M_\theta\equiv\left\lbrack\frac{n_im_iv_{i\theta}^2+n_em_ev_{e\theta}^2}
 {B_\theta^2}\right\rbrack^{1/2}
 $$
 is the poloidal Mach number. Because of  symmetry the toroidal velocity as
 well as the velocity shear (of both toroidal and poloidal components)
 do not appear in (\ref{4a}). It is convenient to use (\ref{4a}) instead of (\ref{3})
 for the electrons. Therefore the slightly reduced  set of equilibrium equations
 we will use in the following consists of  Eqs. (\ref{2}), (\ref{3}), (\ref{3a})
 for ions only, (\ref{5}), (\ref{6}), (\ref{7}), (\ref{8}),  and (\ref{4a}). By
 expressing $B_\theta$ in terms of the safety factor, $q=rB_\theta/(R_0B_z)$,
 with $2\pi R_0$  associated with  the length of the plasma column, and introducing
 the normalized radius $\rho=r/r_0$ with $r_0$ corresponding to the plasma surface,
 Eq. (\ref{4a}) can be put in the form
 \beq
 P'(\rho)=-B_z(\rho)B_z'(\rho)\bigg[1+\Big(\epsilon\frac{\rho}{q(\rho)}\Big)^2\bigg]+
 \left\lbrack M_\theta(\rho)^2+s(\rho)-2\right\rbrack
 \frac{\rho}{r_0}\Big(\epsilon\frac{B_z(\rho)}{q(\rho)}\Big)^2.
 \label{17}
 \eeq
 Here, $\epsilon=r_0/R_0$ is the inverse aspect ratio and
 $s(\rho)=(r/q)(dq/dr)$ the magnetic shear.

\begin{figure}[!ht]
 \begin{center}
 \psfrag{rho}{$\rho$}
 \psfrag{q}{$q(\rho)$}
 \includegraphics[scale=0.8]{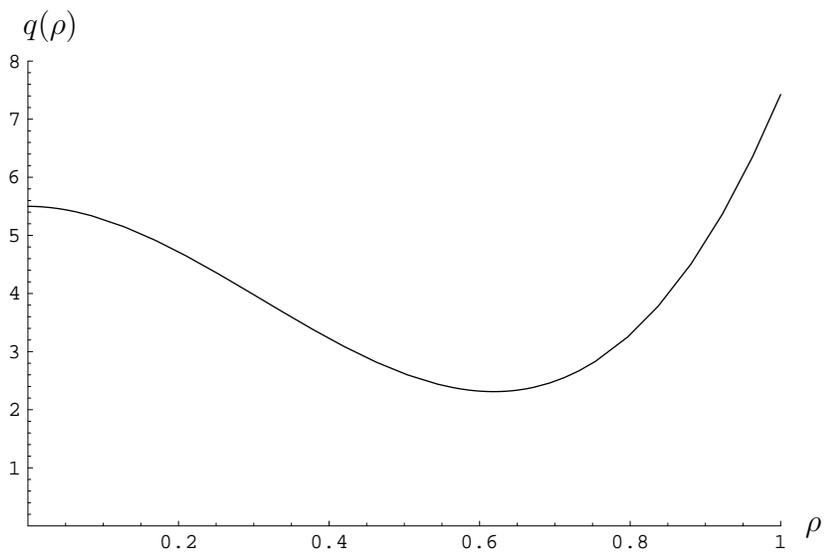}
 \caption{Safety factor profile in connection with Eq.
 (\ref{18}) compatible with the  experimental one measured
 in JT-60U \cite{Koi} (figure 10 therein).}
 \label{fig:1}
 \end{center}
 \end{figure}

 On account of typical experimental ITB profiles we prescribe the
 quantities $q$, $B_z$, $v_{i\theta}$, $v_{iz}$ and $n$ as follows:

\noindent
 Reversed magnetic shear  profile:
 \beq
 q(\rho)=q_c\Big(1-\frac{3\Delta
 q}{q_c}\frac{r_0^2}{r_{min}^2}\rho^2+\frac{2\Delta q}{q_c}
 \frac{r_0^3}{r_{min}^3}\rho^3\Big)
 \label{18}
 \eeq
 where $q_c=q(r=0)$, $r_{min}$ is the position  of $q_{min}$, and
 $\Delta q=q_c-q_{min}$. The shape of the $q$ profile is
 determined by adjusting the parameters $q_{min}$, $\Delta q$ and
 $r_{min}$. Note that $|s|$ is proportional to $\Delta q$; therefore
 as $\Delta q$ takes larger values the magnetic shear increases in
 both the $s<0$ and $s>0$ regions.
 A $q$ profile compatible with experimental ones (see for example figure
 10 in Ref. \cite{Koi}) is presented in figure \ref{fig:1}.

 \noindent
 Toroidal magnetic field profile:
 \beq
 B_z=B_{z0}[1+\delta (1-\rho^2)]^{1/2},
 \label{20}
 \eeq
 where $B_{z0}$ is the vacuum magnetic field and the parameter
 $\delta$ is related to the magnetic properties of the plasma, i.e.
 for $\delta<0$ the plasma is diamagnetic.

 \noindent
 Gaussian-like ion poloidal velocity profile:
 \beq
 v_{i\theta}=4v_{i\theta 0}\rho(1-\rho)\exp{\Big(-\frac{(\rho-\rho_{min})^2}{
 h}\Big)},
 \label{21}
 \eeq
 where the parameter $h>0$
 is related to the
 velocity shear, i.e. $|v_{i\theta}^\prime|$ increases when $h$ takes smaller
 values, and $v_{i\theta 0}$  defines the extremum of $v_{i\theta}$.

 \noindent
 Either peaked on axis toroidal velocity profile:
 \beq
 v_{iz}=v_{iz0}(1-\rho^3)^3
 \label{22}
 \eeq
 or Gaussian-like $v_{iz}$ profile similar to that of (\ref{21});
 it is also noted that the results  do not change if,  alternative
 to (\ref{22}), a peaked on axis toroidal velocity profile of the form
 $$
 v_{iz}=v_{iz0}(1-\rho)\exp \left(-\frac{\rho^2}{h}\right)
 $$
 is employed;

 \noindent
 density profile:
 \beq
 n=n_0(1-\rho^3)^3.
 \label{23}
 \eeq
 In addition, the ion pressure can be expressed in terms of the total
 pressure by the relation
 \beq
 P_i=\lambda P \ \mbox{ , } \ \ 0<\lambda <1 \ .
 \label{24}
 \eeq

 Since in tokamaks $M_\theta<0.1$, the flow term in (\ref{17}) is perturbative
 around the ``static" equilibrium $M_\theta=0$ and therefore it can be
 neglected. It should be noted, however, that this approximation may be not
 good for non-circular cylindrical or axisymmetric plasmas because in these
 cases the convective term in the momentum equation depends on the velocity
 shear which in certain regions may become large (see for example the
 $z$-independent cylindrical and axisymmetric incompressible MHD equilibrium
 equations (23) and (22) in Refs. \cite{ThTa} and \cite{TaTh} respectively).
 The following quantities then can be calculated self-consistently: the poloidal
 magnetic field, $B_\theta=\epsilon\rho B_z/q$, the magnetic shear
 $s=(r/q)(dq/dr)$, the current density via Amp\'ere's law, the
 pressure by integration of (\ref{17}) and setting $P(1)=0$, the ion
 and electron pressures $P_i=\lambda P$ and $P_e=(1-\lambda)P$,  the
 electric field by Eq. (\ref{4}) for the ions
 \beq
 E_r(\rho)= \frac{1}{e r_0n(\rho)}\frac{dP_i(\rho)}{d
 \rho}+ v_{iz}(\rho)B_\theta(\rho)-v_{i\theta}(\rho)B_z(\rho),
 \label{25}
 \eeq
 its shear $|E_r'|$ and $\omega_{E\times B}$ by (\ref{1}).  Also, the electron
 velocity components $v_{ez}$ and $v_{e\theta}$ can be determined by
 the relation ${\bf J}=ne({\bf v}_i-{\bf v}_e)$. It is noted here that the $\nabla P_i$ term
 in (\ref{25}) can be obtained
 in the framework of the ideal Hall-MHD model, alternatively to the complete two fluid one,
 which includes the generalized Ohm's law:
 \begin{equation}
 {\bf E} + {\bf v}  \times {\bf B} = \frac{1}{en}\left( {\bf J}\times {\bf B} -
 \nabla P_e\right).
                                                    \label{25a}
 \end{equation}
Neglecting in the Hall-MHD momentum equation the convective flow
term (which for the case under consideration corresponds to
$M_\theta=0$), the term  $\bf j \times \bf B$ in (\ref{25a}) can be
expressed in terms of the total pressure gradient: $${\bf j} \times
{\bf B} = \nabla P=\nabla(P_i+P_e);$$  then  Eq. (\ref{25a}) leads
to (\ref{25}).
 The  above prescriptions
 and subsequent suggested calculations consists a procedure to solve
 analytically the set of the two-fluid equations. The calculations have been
 performed analytically by developing a programm for symbolic computation \cite{Pou1}
 in connection with \cite{Math}.

 Inspection of (\ref{25}) implies that in addition to the dependence of $E_r$
 and $E_r^\prime$ on the magnetic shear through the $dP_i/d\rho$ term, $s$ is
 involved in the $v_{iz}$ term through the $q$ dependence of $B_\theta$.
 The quantity $\oms$ is stronger affected  by the magnetic shear  because  $s$ is involved
 in  both the $v_{iz}$ and $v_{i\theta}$ terms of $\oms$ [see Eqs. (\ref{26}) and (\ref{27}) in section 3].
 These observations indicate that there is a synergetic contribution of  magnetic shear and flow
 to $E_r$, $E_r'$, and $\oms$. In this report results not obtainable within the framework of MHD will
 mainly be presented in  next section. MHD results were reported in Ref. \cite{Pou}.
 \vspace{4mm} \\

 \noindent
 {\bf 3. Results}\\

 We have set the following values for some of the parameters:
 $B_{z0}=1T$, $\delta=-0.0975$, $Z_i=1$,
 $r_0=1m$, $R_0=3m$, $n_0=5\times 10^{19} part./m^3$, $\lambda=0.6$.
 The choice $q_{min}\geq 2$ was made because according to
 experimental evidence for $q_{min}<2$ strong MHD activity destroys
 confinement possibly due to a double tearing mode \cite{wol}. A
 similar result was found numerically for one-dimensional
 cylindrical equilibria with hollow currents in \cite{KeTa}.
 Moreover in discharges with reversed magnetic shear in JET
 a correlation was found between the formation of ITBs and $q_{min}$
 reaching an integer value (2 or 3) \cite{Jof2}. The impact of
 the magnetic shear  and flow profiles on the equilibrium characteristics
 was examined by varying the parameters  $\Delta q$, $q_{min}$,
 $r_{min}$, $h$, $v_{ iz0}$ and $v_{i\theta 0}$ in the ranges
 (4-14), (2-3), (0.5-0.6), (0.001-0.1), ($10^5$-$10^6$ m$s^{-1}$)
 and ($10^4$-$10^5$ m$s^{-1}$) respectively; consequently $q_c=q_{min}
 +\Delta q$ varies from 6 to 16 and it is guaranteed that $M_\theta^2
 \approx M_z^2$, where $M_z^2=[n(m_iv_{iz}^2+ m_ev_{ez}^2)]/B_z^2$,
 a scaling typical in tokamaks because $B_z\approx 10B_\theta$ and
 $v_{iz}\approx 10 v_{i\theta}$ \cite{mei,rbel}. The impact of the
 variation of  magnetic shear through $\Delta q$ was studied by keeping
 $r_{min}$ and $q_{min}$ constant, while the impact of $r_{min}$
 and $q_{min}$ was examined with  constant $\Delta q$.

\begin{figure}[!ht]
 \begin{center}
 \psfrag{rho}{$\rho$}
 \psfrag{p}{$P(\rho)/P(0)$}
 \psfrag{dqf}{$\Delta q=4$}
 \psfrag{dqft}{$\Delta q=14$}
 \includegraphics[scale=0.8]{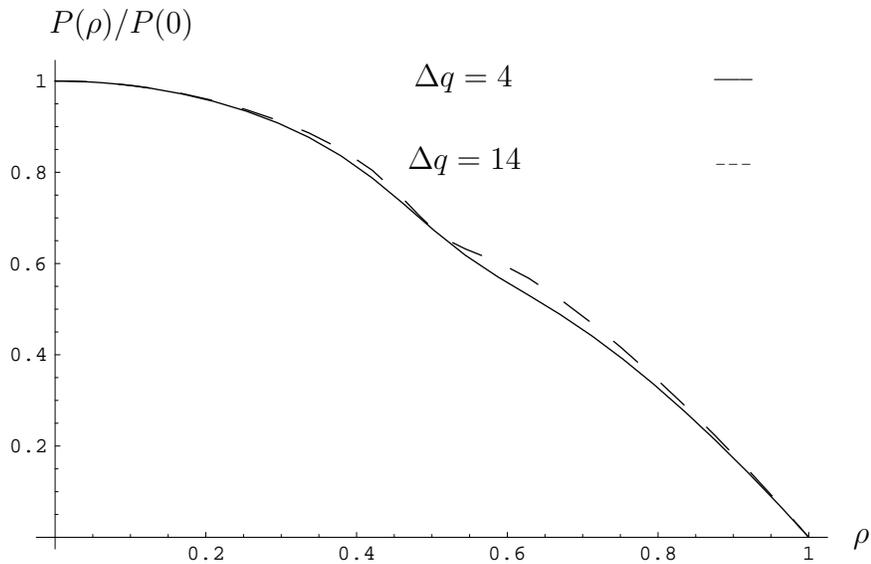}
 \caption{Pressure profiles  for two values of the
 reversed-magnetic-shear parameter $\Delta q$ normalized with respect
 to the value of $P$ at the magnetic axis.}
 \label{fig:2}
 \end{center}
 \end{figure}

\begin{figure}[!ht]
 \begin{center}
 \psfrag{rho}{$\rho$}
 \psfrag{jz}{$J_z(\rho)/J_c$}
 \psfrag{dqf}{$\Delta q=4$}
 \psfrag{dqft}{$\Delta q=14$}
 \includegraphics[scale=0.8]{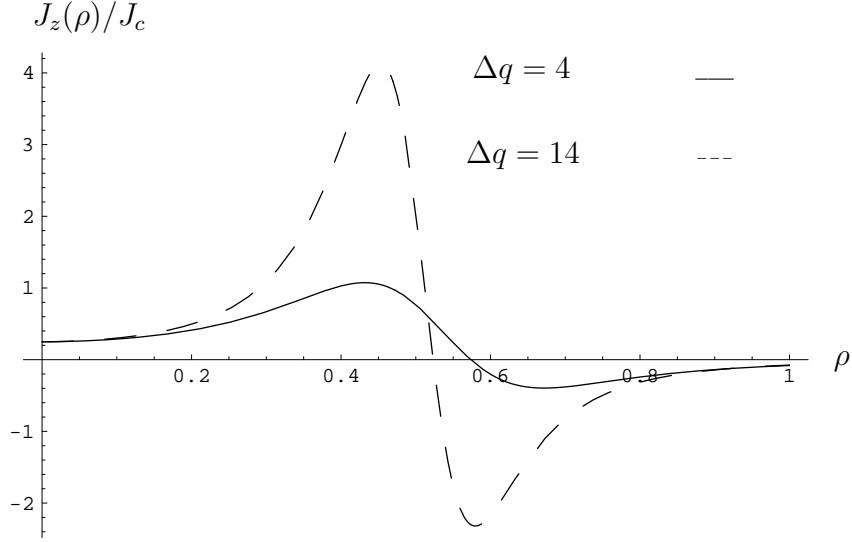}
 \caption{Toroidal current density profiles for two values of $\Delta
 q$ which show the hollow shape and the reversal in the outer
 plasma region. The profiles are normalized with respect to the
 maximum of $J_z$ for $\Delta q=4$.}
 \label{fig:3}
 \end{center}
 \end{figure}

 First we will briefly report certain characteristics of the
 pressure and toroidal current density profiles which remain similar as in
 MHD. The total pressure profile, and therefore the $P_i$ one, is peaked and for $s<0$
 becomes steeper when $|s|$ increases as can be deduced from Eq. (\ref{17})
 (see also figure \ref{fig:2}). In addition (\ref{17}) implies that the
 profile becomes steeper as the plasma becomes more diamagnetic, i.e.
 when $B_z'$ in connection with $\delta$ in (\ref{20}) takes larger values.
 The $J_z$-profile is hollow with its maximum located in the region where
 the $q_{min}$ lies as can be seen in figure \ref{fig:3}. These characteristics
 are observed in discharges with ITBs \cite{Conn} and are favorable for ITB
 formation. Especially for $s>2$ a reversal of $J_z$ occurs in the $s>0$
 region. This characteristic is discussed further   in Ref.  \cite{Pou}. It
 is also noted that a sufficient stability criterion for equilibria with
 reversed current density in the outer plasma area and monotonically
 increasing $q$-profiles was derived in Ref. \cite{LoZa}.

 The conclusions on  the impact of the magnetic shear and flow on $E_r$, $|E_r'|$
 and $\oms$ are reported on an individual basis in the rest of this section. \vspace{2mm} \\

 \noindent
 {\em 3.1 \ Electric field  ($E_r$)}

\begin{figure}[!ht]
 \begin{center}
 \psfrag{rho}{$\rho$}
 \psfrag{er}{$E_r(\rho)/E_c$}
 \psfrag{pr}{$E_{r_{\nabla P_i}}$}
 \psfrag{vz}{$E_{r_{v_{iz}}}$}
 \psfrag{vth}{$E_{r_{v_{i\theta}}}$}
 \includegraphics[scale=0.8]{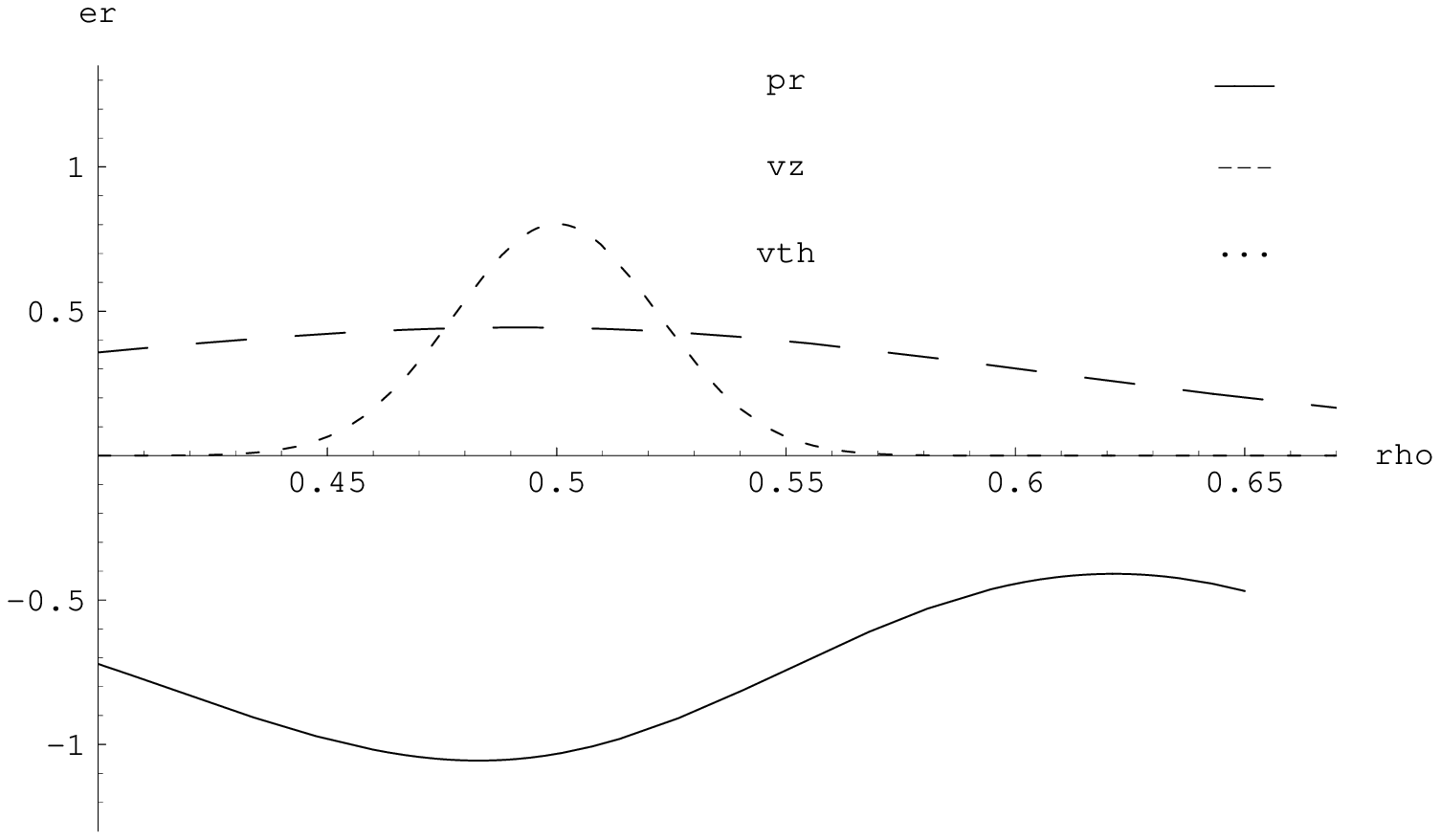}
 \caption{Profiles of the $\nabla P_i$, $v_{iz}$ and $v_{i\theta}$ contributions to
 the electric field, $E_r$, showing that all three contributions are
 of the same order of magnitude. The $v_{iz}$ profile is peaked on
 axis. The profiles are normalized with respect to the extremum of  $\nabla P_i$
 contribution.}
 \label{fig:4}
 \end{center}
 \end{figure}

\begin{figure}[!ht]
 \begin{center}
 \psfrag{rho}{$\rho$}
 \psfrag{er}{$E_r(\rho)/E_c$}
 \psfrag{dqf}{$\Delta q=4$}
 \psfrag{dqft}{$\Delta q=14$}
 \includegraphics[scale=0.8]{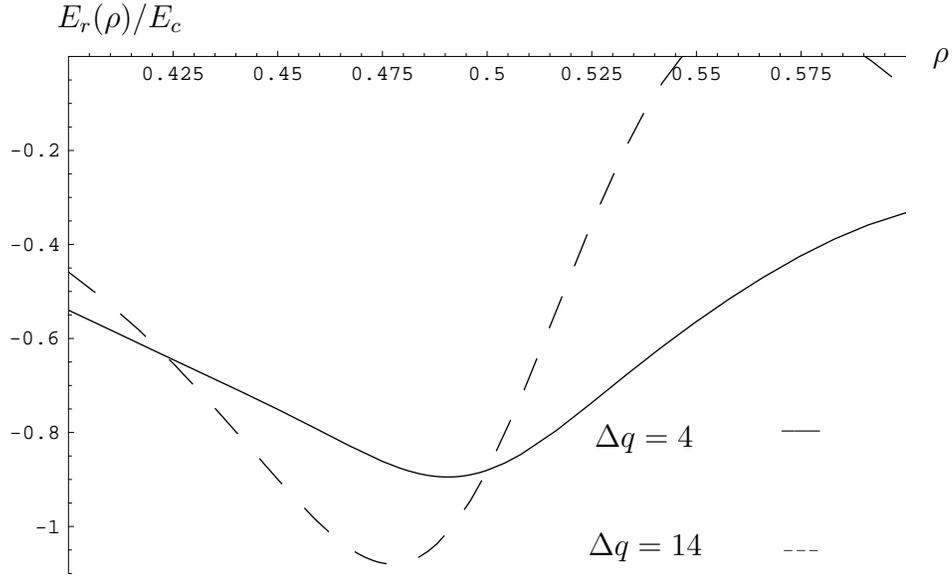}
 \caption{Increase of the normalized absolute value of the electric
 field extremum due to the variation of $\Delta q$ for $q_{min}=4$,
 $r_{min}=0.5$, Gaussian-like $v_{iz}$ profile  and $v_{i\theta}=0$.}
 \label{fig:5}
 \end{center}
 \end{figure}

\begin{figure}[!ht]
 \begin{center}
 \psfrag{rho}{$\rho$}
 \psfrag{er}{$E_r(\rho)/E_c$}
 \psfrag{rf}{$r_{min}=0.5$}
 \psfrag{rs}{$r_{min}=0.6$}
 \includegraphics[scale=0.8]{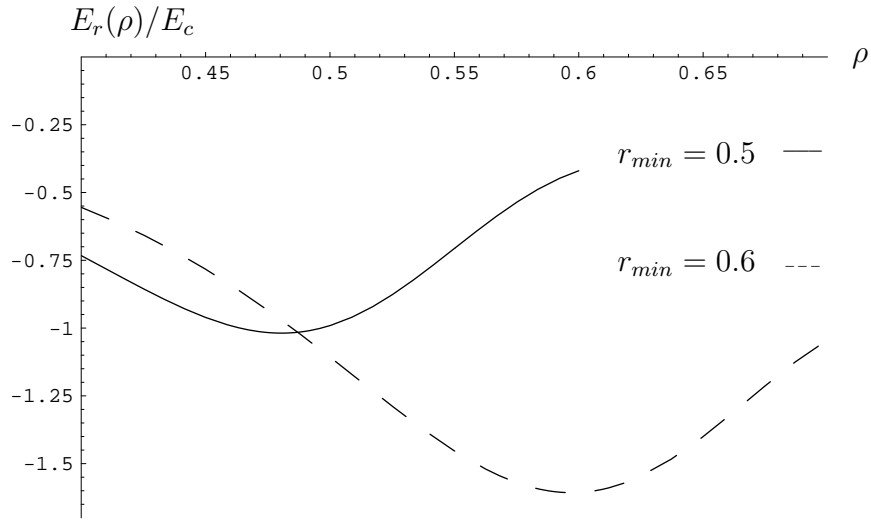}
 \caption{Increase  of the absolute value of the $E_r$
 extremum when the distance $r_{min}$ in connection with the $q_{min}$ position
 becomes larger. Also, the position of the extremum is displaced outwards.
 Here, $q_{min}=2$, $\Delta q=4$, and the flow is purely toroidal peaked on
 axis. The profiles are normalized with respect to the extremum for
 $r_{min}=0.5$.}
 \label{fig:51}
 \end{center}
 \end{figure}

\begin{figure}[!ht]
 \begin{center}
 \psfrag{rho}{$\rho$}
 \psfrag{er}{$E_r(\rho)/E_c$}
 \psfrag{rmt}{$q_{min}=2$}
 \psfrag{rmth}{$q_{min}=3$}
 \includegraphics[scale=0.8]{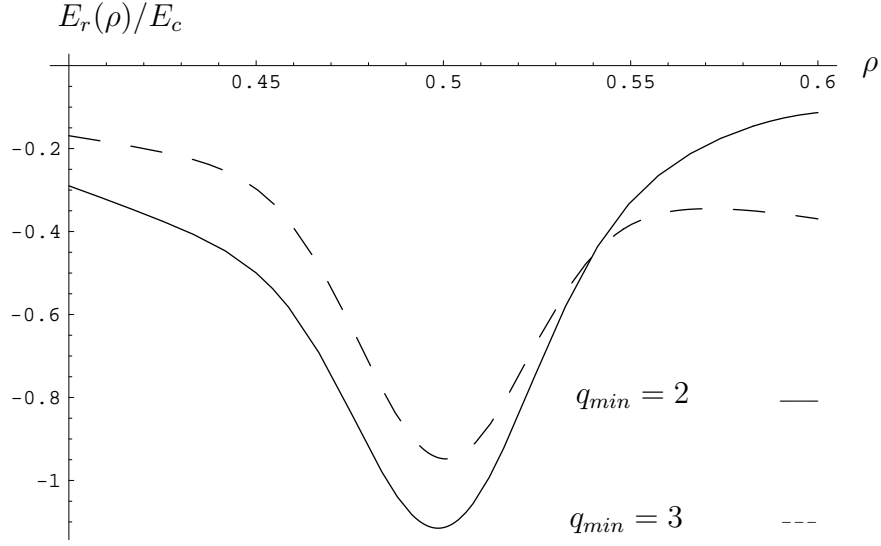}
 \caption{Decrease of the normalized $|E_r|$ maximum when $q_{min}$
 increases for $\Delta q=4$, $r_{min}=0.5$, peaked $v_{iz}$ and
 Gaussian-like localized $v_{i\theta}$ ($h=0.001$).}
 \label{fig:52}
 \end{center}
 \end{figure}

 \begin{enumerate}
 \item
 The electric field  consists of the
 $\nabla P_i$, $v_{iz}$ and, $v_{i\theta}$ contributions in connection with the
 first, second, and  third term in Eq. (\ref{25}).  Each of these terms
 contributes of about the same order of magnitude to $E_r$ (figure \ref{fig:4}).
 This is consistent with experimental evidence \cite{Conn}. A similar result
 was obtained in a different way in \cite{Zhu} (see  figure 4 therein).
 It is apparent from (\ref{25}) that $E_r$ depends linearly on
 $v_{iz}$ and $v_{i\theta}$ with the overall velocity contribution to $E_r$,  however,
 depending on the relative signs of $v_{iz}$, $v_{i\theta}$ and $B_z$.

 \item Typical $E_r$ profiles exhibit an extremum located in the neighborhood
 of the $q_{min}$ position (figure \ref{fig:4}).

 \item Increase of  $|s|$, by increasing $\Delta q$, makes the maximum of
 $|E_r|$ to take larger values (figure \ref{fig:5}). Pending on the direction
 (toroidal or poloidal) of the velocity and the shape of its profile, variation
 of   $\Delta q$ from 4 to 14 increases the values of the $|E_r|$ maximum  in
 a range that varies from $5.6\%$ for purely poloidal flow to $48\%$ for purely
 peaked toroidal flow. It is reminded that, in addition to the $s$ dependence of
 the $\nabla P_i$ term in (\ref{25}), $s$   contributes  to $E_r$ synergetically
 with
 the $v_{iz}$ term (the $v_{i\theta}$ term is $s$ independent).

 \item The larger  $r_{min}$  the higher the values of the $|E_r|$
 maximum (for given values of $\Delta q$ and $q_{min}$)  as shown in
 figure \ref{fig:51}. Quantitatively for a variation of $r_{min}$ from
 0.5 to 0.6, the increase of $|E_r|$ maximum varies from 36\% to 70\%.
 Also  the position of the extremum (located in the vicinity of $r_{min}$)
 is displaced outwards.

 \item The larger  $q_{min}$ the smaller the $|E_r|$ maximum (figure
 \ref{fig:52}). In particular, increase of $q_{min}$ from 2 to 3 (with
 $\Delta q=4$ and $r_{min}=0.5$), results in a decrease of the $|E_r|$
 maximum in the range (12\%, 40\%).

 \item When the  flow shear increases (by decreasing $h$ from 0.1 to
 0.001) the extremum of $E_r$ remains practically unchanged in most
 of the cases considered.
 \end{enumerate}
 \vspace{2mm}

 \noindent
 {\em 3.1 \ Shear of the electric field ($|E_r'|$)}\\

\begin{figure}[!ht]
 \begin{center}
 \psfrag{rho}{$\rho$}
 \psfrag{ser}{$E_r'(\rho)/E_c$}
 \psfrag{pr}{$E_{r_{\nabla P_i}}$}
 \psfrag{vz}{$E_{r_{v_{iz}}}$}
 \psfrag{vth}{$E_{r_{v_{i\theta}}}$}
 \includegraphics[scale=0.8]{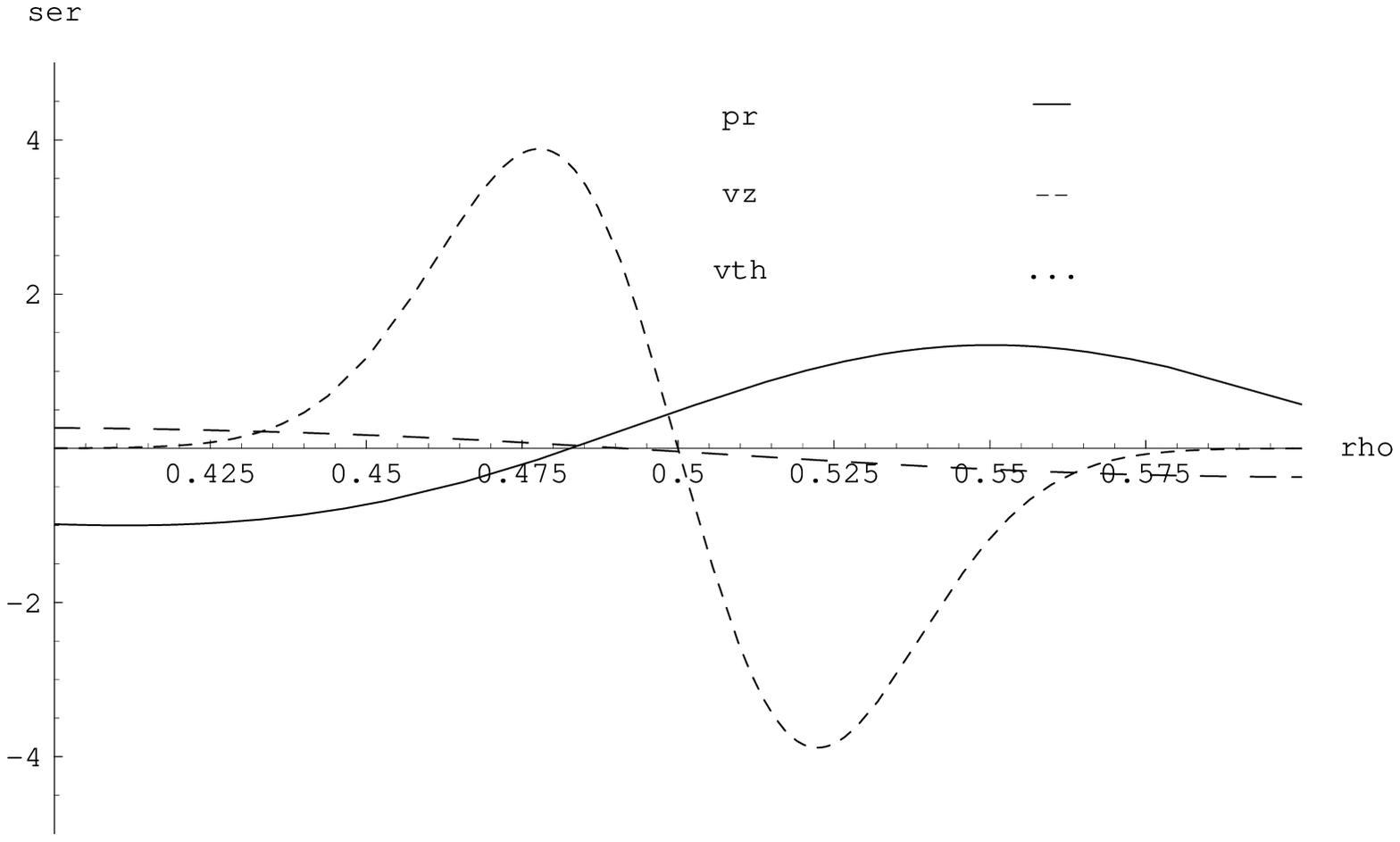}
 \caption{Profiles of the $\nabla P_i$, $v_{iz}$ and $v_{i\theta}$ contributions to
 the electric field shear, $E_r'$, showing that all three contributions are
 of the same order of magnitude. The $v_{iz}$ profile is peaked on
 axis. The profiles are normalized with respect to the extremum of the $\nabla P_i$
 contribution in the $s<0$ region.}
 \label{fig:50}
 \end{center}
 \end{figure}

\begin{figure}[!ht]
 \begin{center}
 \psfrag{rho}{$\rho$}
 \psfrag{ser}{$E_r'(\rho)/E_c'$}
 \psfrag{dqf}{$\Delta q=4$}
 \psfrag{dqft}{$\Delta q=14$}
 \includegraphics[scale=0.8]{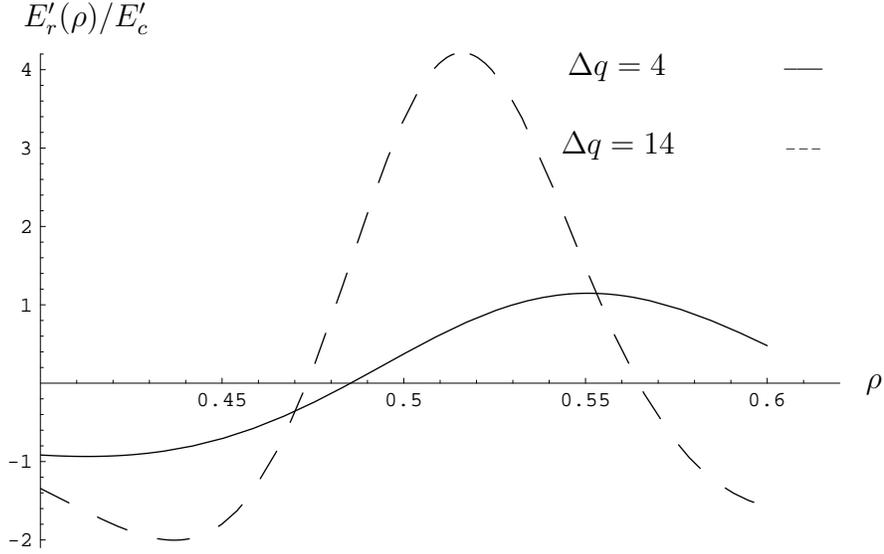}
 \caption{Increase of the normalized $|E_r'|$ maxima caused by an increase
 of the magnetic shear in connection with  variation of  $\Delta q$. The plots
 were obtained for $v_{iz}$ peaked and $v_{i\theta}= 0$.}
 \label{fig:6}
 \end{center}
 \end{figure}

\begin{figure}[!ht]
 \begin{center}
 \psfrag{rho}{$\rho$}
 \psfrag{ser}{$E_r'(\rho)/E'_c$}
 \psfrag{dqf}{$\Delta q=4$}
 \psfrag{dqft}{$\Delta q=14$}
 \includegraphics[scale=0.8]{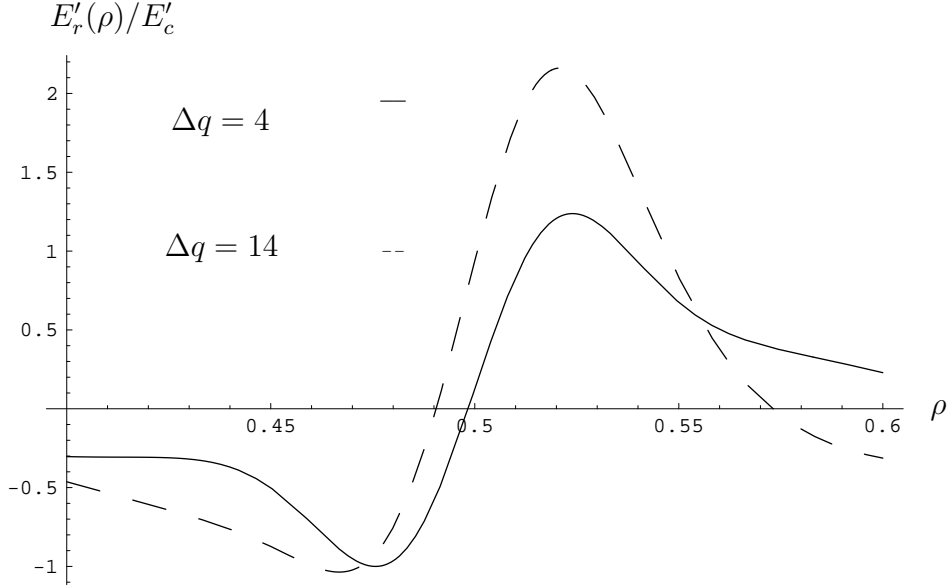}
 \caption{Increase of the $|E_r'|$ extremum in the $s>0$ region due
 to the increase of $\Delta q$. The profiles are obtained for $v_{iz}$
 peaked and $v_{i\theta}\neq 0$ and  are normalized with
 respect to the value of the extremum in the $s<0$ region for $\Delta q=4$.}
 \label{fig:60}
 \end{center}
 \end{figure}

\begin{figure}[!ht]
 \begin{center}
 \psfrag{rho}{$\rho$}
 \psfrag{ser}{$E_r'(\rho)/E_c'$}
 \psfrag{rmf}{$r_{min}=0.5$}
 \psfrag{rms}{$r_{min}=0.6$}
 \includegraphics[scale=0.8]{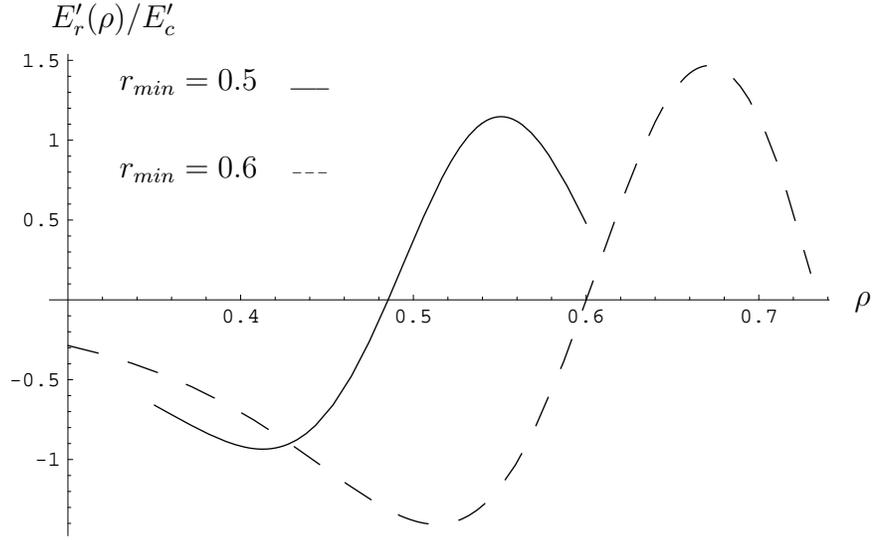}
 \caption{Increase of the
 $|E_r'|$ extrema as $r_{min}$ takes larger values for $q_{min}=2$,
 $\Delta q=4$, and peaked purely  toroidal flow. Also the positions
 of the extrema are displaced outwards. The profiles are normalized with respect to
 the $E_r'$ extremum for $r_{min}=0.5$.}
 \label{fig:61}
 \end{center}
 \end{figure}

\begin{figure}[!ht]
 \begin{center}
 \psfrag{rho}{$\rho$}
 \psfrag{ser}{$E_r'(\rho)/E_c'$}
 \psfrag{qmt}{$q_{min}=2$}
 \psfrag{qmth}{$q_{min}=3$}
 \includegraphics[scale=0.8]{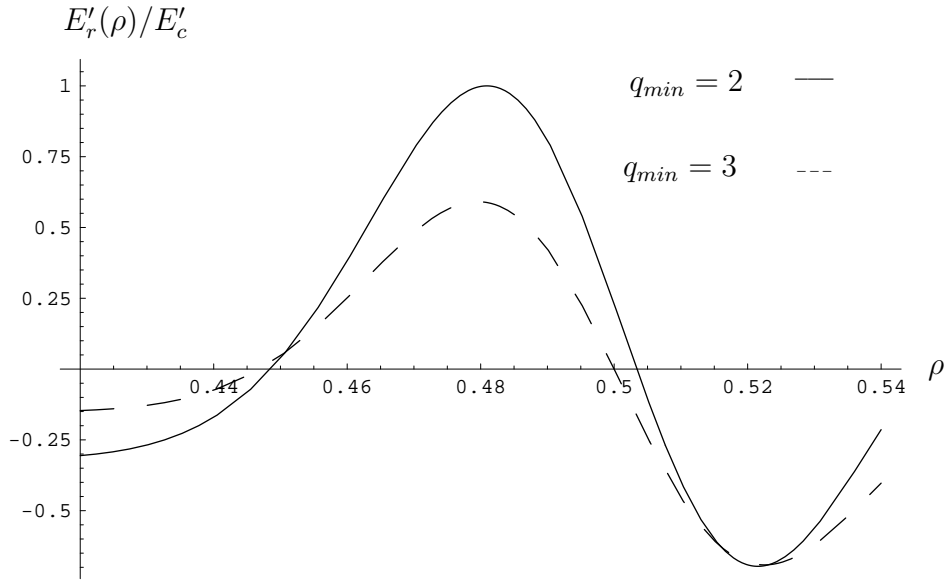}
 \caption{Decrease of the $|E_r'|$ extremum in the $s<0$ region  for
 purely toroidal Gaussian-like flow due to the increase of $q_{min}$.
 For this particular case of flow the variation of the other extremum
 in the $s>0$ region is negligible.}
 \label{fig:62}
 \end{center}
 \end{figure}

\begin{figure}[!ht]
 \begin{center}
 \psfrag{rho}{$\rho$}
 \psfrag{ser}{$E_r'(\rho)/E_c'$}
 \psfrag{hpo}{$h=0.1$}
 \psfrag{hpzzo}{$h=0.001$}
 \includegraphics[scale=0.8]{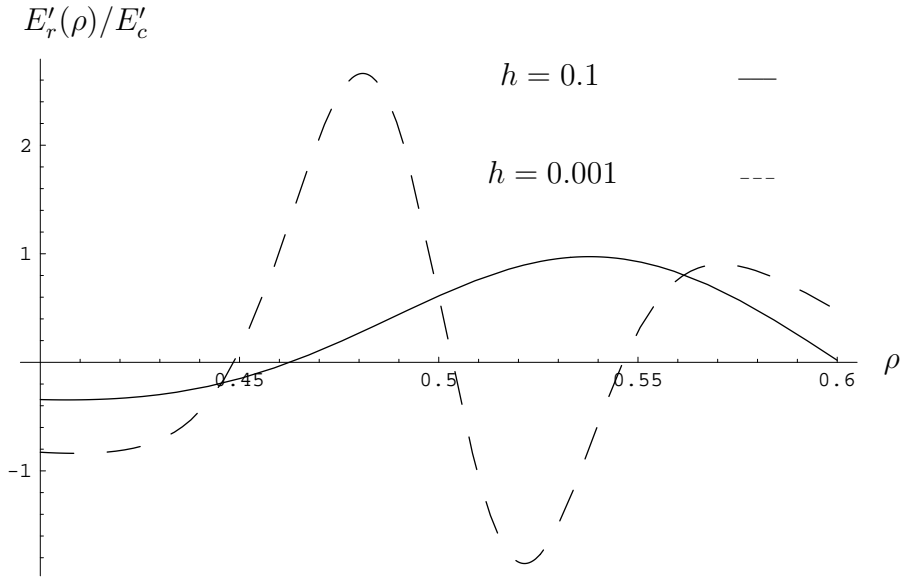}
 \caption{Increase of the $|E_r'|$ extrema due to the increase of the
 flow shear for Gaussian-like $v_{iz}$ and $v_{i\theta}=0$. The
 profiles are normalized with respect to the extremum in the $s>0$
 region for $h=0.001$.}
 \label{fig:7}
 \end{center}
 \end{figure}

 \begin{enumerate}
 \item As in the case of $E_r$ the contributions from the $\nabla P_i$-,
 $v_{iz}$- and $v_{i\theta}$-related terms to $E_r^\prime$ are of
 the same order of magnitude as shown in figure \ref{fig:50}.

 \item The profile of $E_r'$ exhibits one local extremum
 on each side of the $q_{min}$ position (figure \ref{fig:6}). The
 two extrema are of  opposite sign.

 \item Increase of $|s|$ increases both maxima of $|E_r'|$
 in most of the flow cases considered and this increase is
 larger in the region where $s>0$ than that where $s<0$.
 This is shown in figure (\ref{fig:6}). For certain combinations of the
 velocity components, however,  the one extremum increases and the other
 decreases. Such a case with peaked toroidal $v_{iz}$ and poloidal
 $v_{i\theta}$ flow is shown in figure \ref{fig:60}.

 \item The larger $r_{min}$ the higher the maxima of $|E_r'|$  (figure \ref{fig:61})
 unless the case of poloidal velocity in conjunction with
 Gaussian-like toroidal one. Pending on the direction and the shape of
 the velocity  this increase varies from from 8\% to 42\%.
  Also the profile of $|E_r'|$ is displaced outwards as can be seen
  in figure \ref{fig:61}.

 \item  An  increase of  $q_{min}$ results in a decrease of the $|E_r'|$
 extremum in the $s>0$ region in all of the cases considered while in the
 $s<0$ region this happens for $v_{i\theta}=0$ (figure \ref{fig:62}).

 \item By increasing the velocity  shear  the maxima  of $|E_r'|$ are also
 increased in all of the cases considered (figure \ref{fig:7}).

 \item For either purely toroidal or poloidal flow, increase of the
 maximum absolute value of the velocity  by a factor increases  the maxima of  $|E_r'|$
 by the same factor in all of the cases considered with the following
 exception: purely toroidal peaked flow for which the maximum in the
 $s<0$ region increases and the one in the $s>0$ region decreases.

 \item For either purely toroidal or purely poloidal flow, inversion
 of the velocity direction causes a change in the sign of the two
 $E_r'$ extrema. Also, this inversion leads to (i) an increase of both
 maxima of $|E_r'|$ for Gaussian-like $v_{iz}$, (ii) an increase of the
 one $|E_r'|$ maximum in the $s>0$ and a decrease of the other in the $s<0$ region
 and (iii) a decrease of both maxima for purely poloidal flow.
 For Gaussian-like $v_{iz}$ the increase of the one $|E_r'|$-maximum in
 the $s>0$ region caused by inversion is greater than the increase
 of the other in the $s<0$ region.
 \end{enumerate} \vspace{2mm}

 \noindent
 {\em 3.3 \ Shear of the ${\bf E}\times{\bf B}$ velocity ($\oms$)}

\begin{figure}[!ht]
 \begin{center}
 \psfrag{rho}{$\rho$}
 \psfrag{omega}{$\oms(\rho)/\omega_c$}
 \includegraphics[scale=0.8]{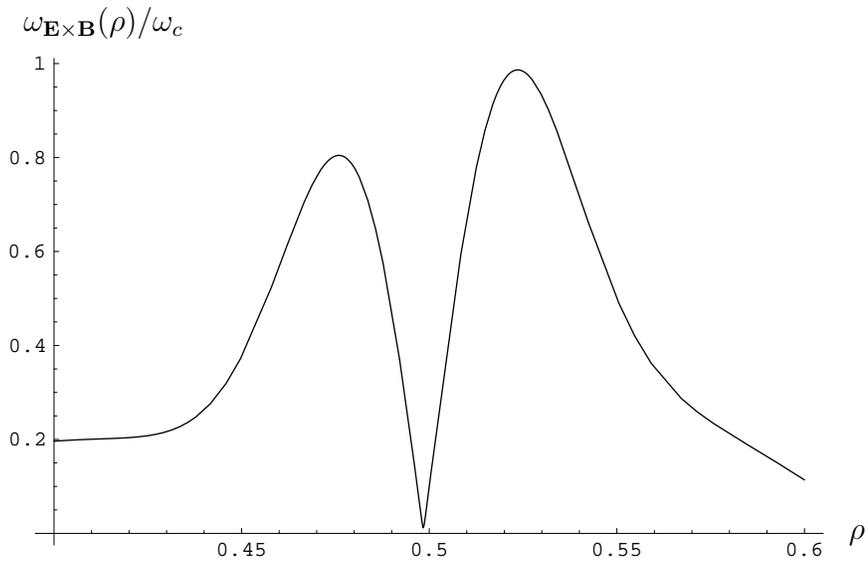}
 \caption{A typical $\oms$ profile for purely poloidal flow,
 normalized with respect to the maximum value in the $s>0$ region.}
 \label{fig:8}
 \end{center}
 \end{figure}

\begin{figure}[!ht]
 \begin{center}
 \psfrag{rho}{$\rho$}
 \psfrag{omega}{$\oms(\rho)/\omega_c$}
 \psfrag{pr}{$E_{r_{\nabla P_i}}$}
 \psfrag{vz}{$E_{r_{v_{iz}}}$}
 \psfrag{vth}{$E_{r_{v_{i\theta}}}$}
 \includegraphics[scale=0.8]{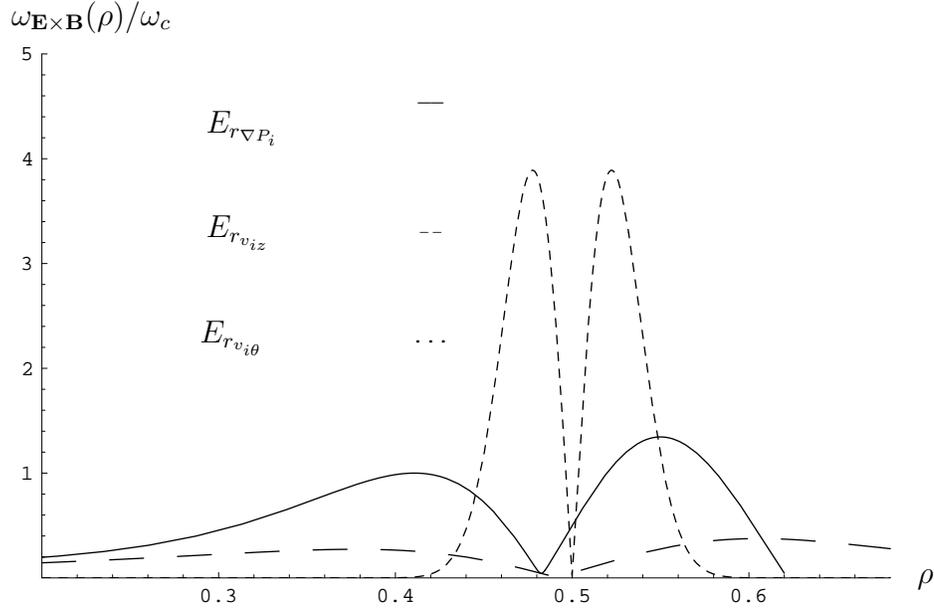}
 \caption{Profiles of the $\nabla P_i$, $v_{iz}$ and
 $v_{i\theta}$ contributions to $\oms$ showing that all three
 are of the same order of magnitude. The $v_{iz}$ profile is peaked on axis.
 The normalization is made  with respect to the  maximum value of the $\nabla
 P_i$ contribution in the $s<0$ region.}
 \label{fig:81}
 \end{center}
 \end{figure}

 \begin{enumerate}
 \item The profile of $\oms$ (Eq. \ref{1})
 possesses two maxima located the one in the $s<0$ and the other in the
 $s>0$ regions (figure \ref{fig:8}). Larger of the two maxima is the
 one which lies in the region of steeper pressure profile.

 \item The impact of the magnetic shear on  $\omega_{{\bf E}\times{\bf
 B}}$ is stronger than that in  MHD due to the $\nabla P_i$ term of the
 electric field [Eq. (\ref{25})]. Specifically for constant $B_z$ and
 arbitrary profiles of $q$, $v_{iz}$ and $v_{i\theta}$,  Eq. (\ref{1})
 yields at the point where $E_r'=0$
 \beq
 \omega_{{\bf E}\times{\bf
 B}}= \Big|\lambda\frac{(1-s)(2-s)B_z\rho\epsilon}{enqr_0^2(\rho^2+\frac{q^2}{\epsilon^2})}
 -\omega_{{\bf E}\times{\bf B}-MHD}\Big|,
 \label{26}
 \eeq
 where
 \beq
 \omega_{{\bf E}\times{\bf
 B}-MHD}=\frac{(1-s)\big(\frac{\epsilon\rho v_{iz}}{q}-v_{i\theta}\big)}{
 r_0^2\big(\rho^2+\frac{q^2}{\epsilon^2}\big)}
 \label{27}
 \eeq
 The first term in (\ref{26}) stems from the $\nabla P_i$ part of
 $E_r$ in (\ref{17}) while the second term comes from the $v_{iz}$
 and $v_{i\theta}$ parts of $E_r$. The subscript MHD is used to
 emphasize the similarity of (\ref{27}) with the respective MHD
 relation derived in Ref. \cite{Pou} [equation (18) therein]. It is
 apparent the $\nabla P_i$-related dependence of $\omega_{{\bf E}
 \times{\bf B}}$ on $s$, proportional to $(1-s)(2-s)$,  is stronger
 than the $v_{iz}$ and $v_{i\theta}$ dependence proportional to $1-s$;
 also, the absolute values of the $\nabla P_i$-, $v_{iz}$- and
 $v_{i\theta}$- related terms are individually larger for $s<0$ than $s>0$.
 The contribution of each of these terms to $\oms$ however  is of the
 same order of magnitude. Note that despite of the tokamak pertinent scaling
 $v_{iz}\approx 10 v_{i\theta}$, the contributions of $v_{iz}$ and
 $v_{i\theta}$ terms  are of the same order of magnitude because
 of the factor $\epsilon \rho/q$. The ``equipartition" of the three
 terms holds in general for the whole $\oms$ profile obtained via
 the symbolic computation programme as shown in Fig \ref{fig:81}.

 \item Increase of the flow via  either $|v_{iz0}|$ or  $|v_{i\theta 0}|$
 by a factor increases the maxima of $\oms$ by the same factor.
 \end{enumerate}

\begin{figure}[!ht]
 \begin{center}
 \psfrag{rho}{$\rho$}
 \psfrag{omega}{$\oms(\rho)/\omega_c$}
 \psfrag{dqf}{$\Delta q=4$}
 \psfrag{dqft}{$\Delta q=14$}
 \includegraphics[scale=0.8]{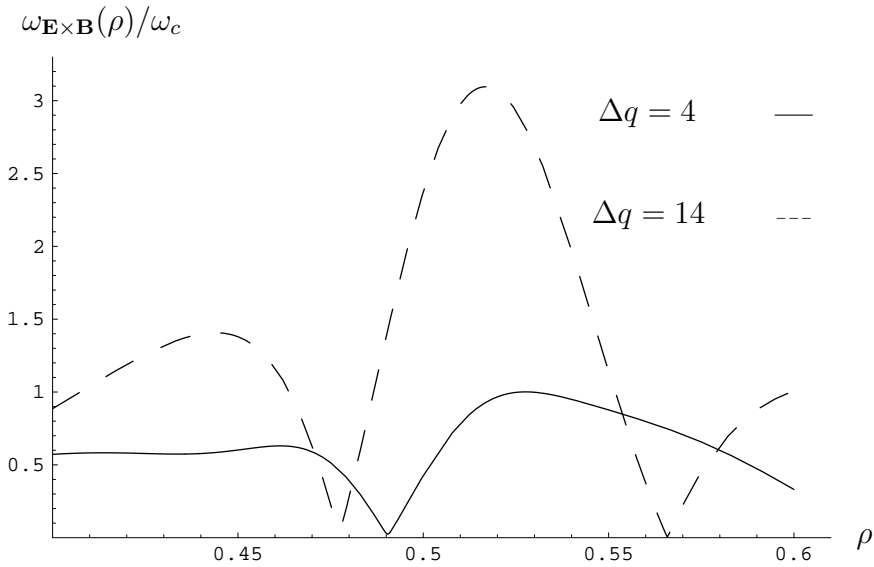}
 \caption{Increase of the normalized $\oms$ maxima due to the
 increase of the magnetic shear. Both velocity components have
 Gaussian-like profiles.}
 \label{fig:9}
 \end{center}
 \end{figure}

\begin{figure}[!ht]
 \begin{center}
 \psfrag{rho}{$\rho$}
 \psfrag{omega}{$\oms(\rho)/\omega_c$}
 \psfrag{dqf}{$\Delta q=4$}
 \psfrag{dqft}{$\Delta q=14$}
 \includegraphics[scale=0.8]{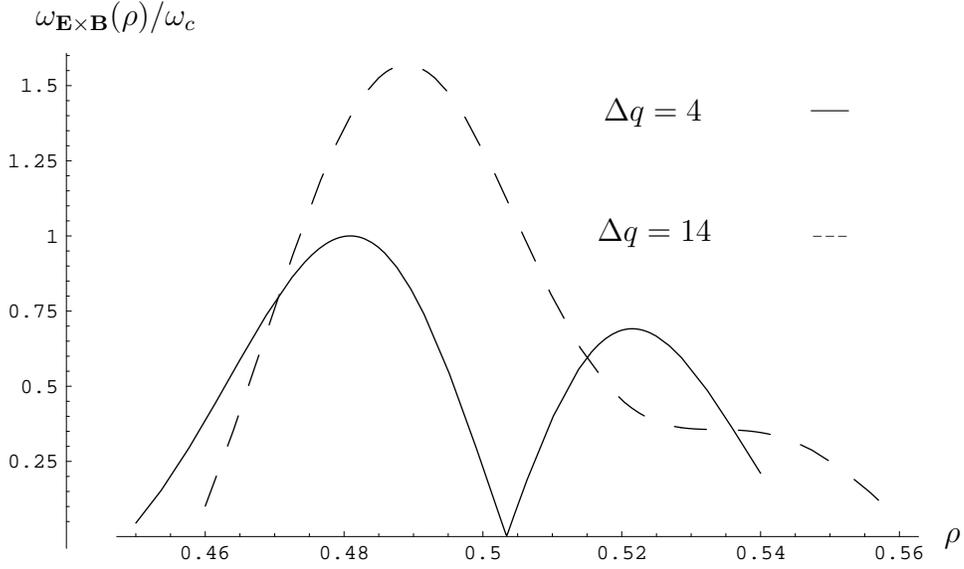}
 \caption{Increase of the $\oms$ extremum in the $s<0$ and decrease in
 the $s>0$ regions due to the increase of $\Delta q$. The profiles are
 obtained for $v_{iz}$ Gaussian-like and $v_{i\theta}=0$ and are
 normalized with respect to the $s<0$ extremum of $\oms$ for $\Delta q=4$.}
 \label{fig:90}
 \end{center}
 \end{figure}

\begin{figure}[!ht]
 \begin{center}
 \psfrag{rho}{$\rho$}
 \psfrag{omega}{$\oms(\rho)/\omega_c$}
 \psfrag{rmf}{$r_{min}=0.5$}
 \psfrag{rms}{$r_{min}=0.6$}
 \includegraphics[scale=0.8]{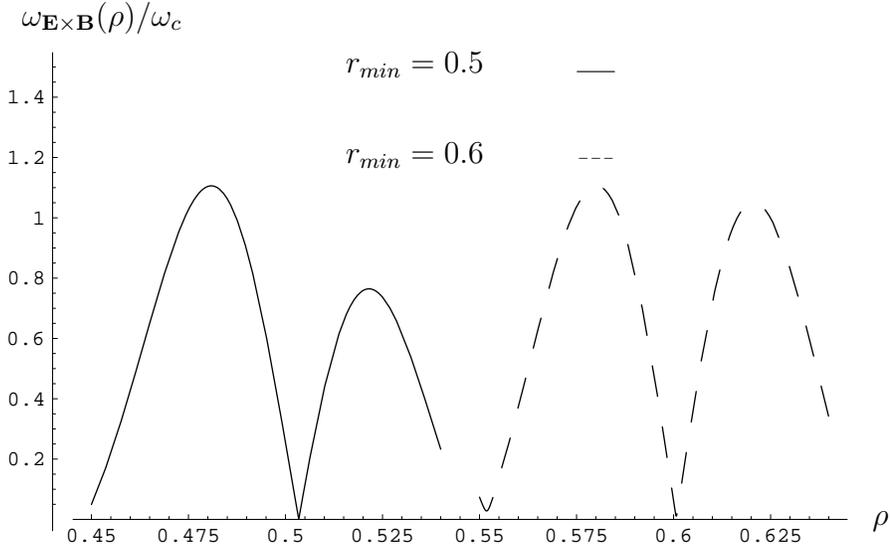}
 \caption{Increase of the normalized $\oms$ maxima due to the
 outward shift of the position of $q_{min}$ for purely toroidal
 Gaussian-like flow.}
 \label{fig:91}
 \end{center}
 \end{figure}

\begin{figure}[!ht]
 \begin{center}
 \psfrag{rho}{$\rho$}
 \psfrag{omega}{$\oms(\rho)/\omega_c$}
 \psfrag{qmt}{$q_{min}=2$}
 \psfrag{qmth}{$q_{min}=3$}
 \includegraphics[scale=0.7]{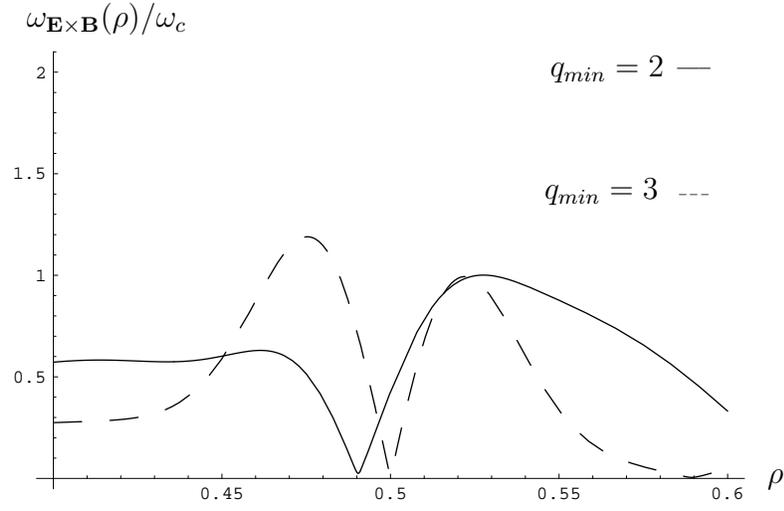}
 \caption{Increase of the $\oms$-extrema as $q_{min}$ takes larger
 values when  both velocity components have Gaussian-like profiles.
 In this particular case the increase of the extremum in the $s>0$ region is very
 small.  The profiles are normalized with respect to the $s>0$ maximum of $\oms$ for
 $q_{min}=3$.}
 \label{fig:92}
 \end{center}
 \end{figure}

\begin{figure}[!ht]
 \begin{center}
 \psfrag{rho}{$\rho$}
 \psfrag{omega}{$\oms(\rho)/\omega_c$}
 \psfrag{hpo}{$h=0.1$}
 \psfrag{hpzzo}{$h=0.001$}
 \includegraphics[scale=0.8]{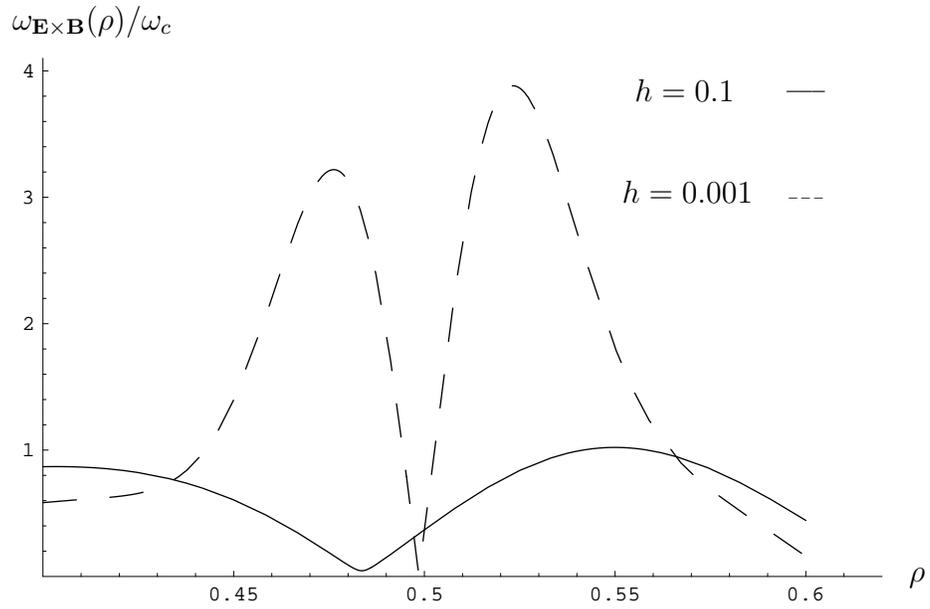}
 \caption{Increase of the $\oms$ extrema
 caused by the increase of the flow shear for $v_{iz}$ peaked
 and $v_{i\theta}\neq 0$.  The normalization is made with respect to the $\oms$
 extremum in the
 $s>0$ region  for $h=0.1$.}
 \label{fig:10}
 \end{center}
 \end{figure}

\begin{figure}[!ht]
 \begin{center}
 \psfrag{rho}{$\rho$}
 \psfrag{omega}{$\oms(\rho)/\omega_c$}
 \psfrag{vzp}{$v_{i z}>0$}
 \psfrag{rms}{$v_{i z}<0$}
 \includegraphics[scale=0.7]{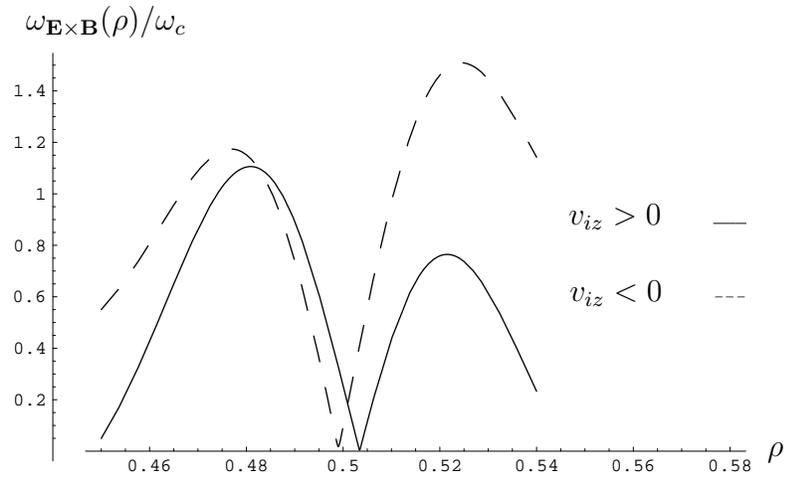}
 \caption{Increase of the normalized $\oms$-extrema caused by inversion of a
 Gaussian-like toroidal velocity.}
 \label{fig:101}
 \end{center}
 \end{figure}

 The impact of the magnetic shear through $\Delta q$ and the flow on
 $\oms$ is similar as that on $E_r^\prime$. Specifically:

 \begin{enumerate}
 \item Increase of $|s|$  leads to larger values for the maxima of
 $\omega_{{\bf E}\times{\bf B}}$ in most of the flow cases considered
 (figure \ref{fig:9}). There are some combinations of velocity components
 however for which the one maximum increases and the other decreases. Such
 a case is shown in figure \ref{fig:90}. In which region ($s<0$ or $s<0$) the
 increase takes place depends on the particular velocity components
 involved and the shape of the toroidal velocity profile.

 \item
 The larger $r_{min}$ the greater the $\oms$-maxima (figure \ref{fig:91})
 in the same cases as for $E_r'$. The profile of $\oms$ is also displaced
 outwards as can be seen in figure \ref{fig:91}.

 \item Increase of $q_{min}$ causes (i) a decrease of the $\omega_{{\bf
 E}\times{\bf B}}$-maximum in the $s>0$ region in all of the cases
 considered and (ii) an increase of  $\omega_{{\bf E}\times{\bf B}}$
 -maximum in the $s<0$ one for $v_{i\theta}\neq0$ (figure \ref{fig:92}).

 \item The larger the flow shear the greater the $\oms$ maxima in all
 of the cases considered (figure \ref{fig:10}).

 \item $\oms$ is affected by the relative signs of $v_{i\theta}$,
 $v_{iz}$ and $B_z$ as is apparent from Eqs. (\ref{26}) and (\ref{27}).
 In particular (i) inversion of the Gaussian-like toroidal velocity
 increases the maxima of $\oms$ (figure \ref{fig:101}), (ii) the maximum
 of $\oms$ in the $s>0$ region increases while the one in the $s<0$ region
 decreases due to the reversal of the peaked toroidal velocity and (iii)
 they decrease by inversion of the poloidal velocity. Finally for both
 Gaussian-like velocity components the variation caused by inversion is
 greater in the $s>0$ region.
 \end{enumerate} \vspace{4mm}

 \noindent
 {\bf 4. Summary and Conclusions}\\

 In this report tokamak equilibria with reversed magnetic shear and sheared
 flow have been studied within the framework of two-fluid model in the limit
 of infinite aspect ratio. The study is based on a slightly reduced set of
 two-fluid equations in which the electron  momentum equation is replaced by
 the respective MHD one. Neglecting the flow term in this equation (because
 in cylindrical geometry it is small for tokamaks) and  prescribing the
 profiles of six free quantities in accord with ITB experimental ones, i.e.
 the toroidal magnetic field $B_z$, the safety factor $q$, the toroidal and
 poloidal ion velocities $v_{iz}$ and $v_{i\theta}$, the density $n$
 [Eqs. (\ref{20})-(\ref{23})]
 and the ion pressure in terms of the total pressure, $P_i=\lambda P$, we have
 constructed analytic solutions in
 calculating self consistently the following quantities:
 $P$ [and therefore $P_i$ and the  electron pressure
 $P_e=(1-\lambda)P$], the  current density and the radial electric field
 $E_r$; the electric field shear,   $|E_r'|$, and the shear of the ${\bf E}
 \times {\bf B}$-velocity, $\oms$ [Eq. (\ref{1})] have also been calculated.
 Gaussian-like profiles for $v_{i\theta}$ and either Gaussian-like or
 peaked-on-axis ones for $v_{iz}$ have been considered. In addition, for
 reversed magnetic shear profiles the impact of $s$  and the flow on the
 equilibrium characteristics  has been examined  by varying the parameters
 $\Delta q$ which $|s|$ is proportional to, the minimum of $q$, $q_{min}$,
 its position, $r_{min}$, the extrema of the velocity components, $v_{iz0}$
 and $v_{i\theta0}$, and a parameter $h$ which decreases with increasing velocity
 shear. The results  are as follows.

 \begin{enumerate}
 \item The pressure profiles become steeper in the region of $s<0$.

 \item The profile of the toroidal current density $J_z$ is hollow and a
 reversal occurs in the outer plasma region for $s>2$ in connection with
 appropriate values of $\Delta q$.

 \item The $|E_r|$ profile has a maximum located close to the $q_{min}$
 position while the $|E_r'|$ and $\oms$ ones have two local maxima the
 one in the $s>0$ and the other in the $s<0$ regions.

 \item The contributions associated with  $\nabla P_i$,  $v_{iz}$, and $v_{i\theta}$
 to $E_r$, $E_r'$ and $\oms$ (the $\nabla P_i$ contribution being
 missed in MHD)
 are of the same order of magnitude.

 \item The magnetic shear affects $E_r$ and $E_r'$ explicitly through
 $\nabla P_i$ and implicitly in conjunction with $v_{iz}$; $s$ has an
 additional impact on $\oms$  in connection with $v_{i\theta}$. The
 explicit impact of $s$ is stronger; in particular for $B_z=$constant,
 the $\nabla P_i$ contribution to $\oms$ at the point where $E_r'=0$
 is proportional to $(1-s)(2-s)$ [Eq. (\ref{26})] while the contribution
 through the flow terms is proportional to (1-s) [Eq. (\ref{27})].

 \item  Increase of $|s|$ results in an increase in the maximum of
 $|E_r|$ in all of the cases considered. Also, the maxima of $|E_r'|$ and $\oms$
 increase  in  most of the flow cases considered. When either the toroidal and poloidal
 velocity contributions cancel each other or the velocity is purely
 toroidal peaked,
   the
 increase is greater in the
 $s>0$ region. Also
 pending on the direction and shape of the flow, the increase
 varies
 from $56.4\%$ to $323\%$.

 \item The larger $r_{min}$ the greater the maxima of $|E_r|$,
 $|E_r'|$, and $\oms$.

 \item The larger $q_{min}$ the smaller the maximum of $|E_r|$  but the
 larger the maxima of $|E_r'|$ and $\oms$  in the $s>0$ region.

 \item  Stronger flow, by  larger values of $|v_{iz0}|$ and
 $|v_{i\theta0}|$, leads to linear increase in $E_r$, $|E_r'|$, and $\oms$.

 \item The larger the flow shear (by smaller values of  the parameter h) the
 slightly smaller the maximum of $|E_r|$ but the larger the maxima of $|E_r'|$
 and $\oms$.

 \item $E_r$, $E_r'$, and $\oms$ are sensitive to the relative signs
 of $v_{iz}$, $v_{i\theta}$, and $B_z$.
 \end{enumerate}

In summary, alike MHD, in the framework of two-fluid model the
magnetic shear and  sheared flow (toroidal and poloidal) act
synergetically on $E_r$, $E_r'$, and $\oms$ which may play a role in
the formation of ITBs.  However the impact of magnetic shear on
these quantities is stronger than that in MHD due to the additional
$\nabla P_i$ contribution to the aforementioned terms.

\begin{center}
 {\large\bf Acknowledgements}
 \end{center}

 Part of this work was conducted during a visit of the authors G.P.
 and G.N.T. to the Max-Planck-Institut f\"{u}r Plasmaphysik,
 Garching. The hospitality of that Institute is greatly appreciated.

 This  work was performed under the Contract of Association
 ERB 5005 CT 99 0100 between the European Atomic Energy Community and
 the Hellenic Republic.

 \newpage

 \begin{center}
 {\large \bf  Figure captions}
 \end{center}

 \noindent
 Figure \ref{fig:1}: \ Safety factor profile in connection with Eq.
 (\ref{18}) compatible with the  experimental one measured
 in JT-60U \cite{Koi} (figure 10 therein).
 \vspace{0.3cm}

 \noindent
 Figure\ref{fig:2}: \ Pressure profiles  for two values of the
 reversed-magnetic-shear parameter $\Delta q$ normalized with respect
 to the value of $P$ at the magnetic axis.
 \vspace{0.3cm}

 \noindent
 Figure  \ref{fig:3}: \ Toroidal current density profiles for two
 values of $\Delta q$ which show the hollow shape and the reversal
 in the outer plasma region. The profiles are normalized with respect
 to the maximum of $J_z$ for $\Delta q=4$.
 \vspace{0.3cm}

 \noindent
 Figure \ref{fig:4}: \ Profiles of the $\nabla P_i$, $v_{iz}$ and $v_{i\theta}$ contributions to
 the electric field, $E_r$, showing that all three contributions are
 of the same order of magnitude. The $v_{iz}$ profile is peaked on
 axis. The profiles are normalized with respect to the extremum of  $\nabla P_i$
 contribution.
 \vspace{0.3cm}

 \noindent
 Figure  \ref{fig:5}: \ Increase of the normalized absolute value of the electric
 field extremum due to the variation of $\Delta q$ for $q_{min}=4$,
 $r_{min}=0.5$, Gaussian-like $v_{iz}$ profile  and $v_{i\theta}=0$.
 \vspace{0.3cm}

 \noindent
 Figure  \ref{fig:51}: \ Increase  of the absolute value of the $E_r$
 extremum when the distance $r_{min}$ in connection with the $q_{min}$ position
 becomes larger. Also, the position of the extremum is displaced outwards.
 Here, $q_{min}=2$, $\Delta q=4$, and the flow is purely toroidal peaked on
 axis. The profiles are normalized with respect to the extremum for
 $r_{min}=0.5$.
 \vspace{0.3cm}

 \noindent
 Figure  \ref{fig:52}: \ Decrease of the normalized $|E_r|$ maximum when $q_{min}$
 increases for $\Delta q=4$, $r_{min}=0.5$, peaked $v_{iz}$ and
 Gaussian-like localized $v_{i\theta}$ ($h=0.001$).
 \vspace{0.3cm}

 \noindent
 Figure  \ref{fig:50}: \ Profiles of the $\nabla P_i$, $v_{iz}$ and $v_{i\theta}$ contributions to
 the electric field shear, $E_r'$, showing that all three contributions are
 of the same order of magnitude. The $v_{iz}$ profile is peaked on
 axis. The profiles are normalized with respect to the extremum of the $\nabla P_i$
 contribution in the $s<0$ region.
 \vspace{0.3cm}

 \noindent
 Figure  \ref{fig:6}: \ Increase of the normalized $|E_r'|$ maxima
 caused by an increase of the magnetic shear in connection with
 variation of  $\Delta q$. The plots were obtained for $v_{iz}$
 peaked and $v_{i\theta}= 0$.
 \vspace{0.3cm}

  \noindent
 Figure  \ref{fig:60}: \  Increase of the $|E_r'|$ extremum in the $s>0$ region due
 to the increase of $\Delta q$. The profiles are obtained for $v_{iz}$
 peaked and $v_{i\theta}\neq 0$ and  are normalized with
 respect to the value of the extremum in the $s<0$ region for $\Delta q=4$.
 \vspace{0.3cm}

 \noindent
 Figure  \ref{fig:61}: \  Increase of the
 $|E_r'|$ extrema as $r_{min}$ takes larger values for $q_{min}=2$,
 $\Delta q=4$, and peaked purely  toroidal flow. Also the positions
 of the extrema are displaced outwards. The profiles are normalized with respect to
 the $E_r'$ extremum for $r_{min}=0.5$
 \vspace{0.3cm}

 \noindent
 Figure \ref{fig:62}: \ Decrease of the $|E_r'|$ extremum in the $s<0$ region  for
 purely toroidal Gaussian-like flow due to the increase of $q_{min}$.
 For this particular case of flow the variation of the other extremum
 in the $s>0$ region is negligible.
 \vspace{0.3cm}

 \noindent
 Figure \ref{fig:7}: \ Increase of the $|E_r'|$ extrema due to
 the increase of the flow shear for Gaussian-like $v_{iz}$ and $v_{i\theta}=0$.
 \vspace{0.3cm}

 \noindent
 Figure \ref{fig:8}: \ A typical $\oms$ profile for purely poloidal flow,
 normalized with respect to the maximum value in the $s>0$ region.
 \vspace{0.3cm}

 \noindent
 Figure  \ref{fig:81}: \ Profiles of the $\nabla P_i$, $v_{iz}$ and
 $v_{i\theta}$ contributions to $\oms$ showing that all three
 are of the same order of magnitude. The $v_{iz}$ profile is peaked on axis.
 The normalization is made  with respect to the  maximum value of the $\nabla
 P_i$ contribution in the $s<0$ region.
 \vspace{0.3cm}

 \noindent
 Figure \ref{fig:9}: \ Increase of the normalized $\oms$ maxima due to the
 increase of the magnetic shear. Both velocity components have
 Gaussian-like profiles.
 \vspace{0.3cm}

  \noindent
 Figure \ref{fig:90}: \ Increase of the $\oms$ extremum in the $s<0$ and decrease in
 the $s>0$ regions due to the increase of $\Delta q$. The profiles are
 obtained for $v_{iz}$ Gaussian-like and $v_{i\theta}=0$ and are
 normalized with respect to the $s<0$ extremum of $\oms$ for $\Delta q=4$.
 \vspace{0.3cm}

 \noindent
 Figure \ref{fig:91}: \ Increase of the normalized $\oms$ maxima due to the
 outward shift of the position of $q_{min}$ for purely toroidal
 Gaussian-like flow.
 \vspace{0.3cm}

 \noindent
 Figure \ref{fig:92}: \ Increase of the $\oms$-extrema as $q_{min}$ takes larger
 values when  both velocity components have Gaussian-like profiles.
 In this particular case the increase of the extremum in the $s>0$ region is very
 small.  The profiles are normalized with respect to the $s>0$ maximum of $\oms$ for
 $q_{min}=3$.
 \vspace{0.3cm}

 \noindent
 Figure \ref{fig:10}: \ Increase of the $\oms$ extrema
 caused by the increase of the flow shear for $v_{iz}$ peaked
 and $v_{i\theta}\neq 0$.  The normalization is made with respect to the $\oms$
 extremum in the
 $s>0$ region  for $h=0.1$.
 \vspace{0.3cm}

 \noindent
 Figure \ref{fig:101}: \ Increase of the normalized $\oms$-extrema caused
 by inversion of a Gaussian-like toroidal velocity.

  \newpage
 \begin{center}
 {\large \bf  List of Figures}
 \end{center}

 \clearpage


 \clearpage


 \clearpage


 \end{document}